\documentclass[aps,prc,superscriptaddress,showpacs,twocolumn,floatfix]{revtex4}
\usepackage{epsfig,longtable}

\usepackage{graphicx}

\newcommand{\etal}{{\it et al.}}
\def\gtorder{\mathrel{\raise.3ex\hbox{$>$}\mkern-14mu
             \lower0.6ex\hbox{$\sim$}}}
\def\ltorder{\mathrel{\raise.3ex\hbox{$<$}\mkern-14mu
             \lower0.6ex\hbox{$\sim$}}}

\begin{document}

%%%%%%%%%%%%%%%%%%%%%%%%%%%%%%%%%%%%%%%%%%%%%%%%%%%%%%%%%%%%%%%%%%%%%%%%%%%%
\title{Moments of nuclear and nucleon structure functions at low $Q^2$
  and the momentum sum rule}
%%%%%%%%%%%%%%%%%%%%%%%%%%%%%%%%%%%%%%%%%%%%%%%%%%%%%%%%%%%%%%%%%%%%%%%%%%%%

\author{I. Niculescu}
\affiliation{James Madison University, Harrisonburg, VA 22807, USA}
\author{J. Arrington}
\affiliation{Physics Division, Argonne National Laboratory, Argonne, IL 60439, USA}
\author{R. Ent}
\affiliation{Thomas Jefferson National Accelerator Facility, Newport News, VA 23602, USA}
\author{C. E. Keppel}
\affiliation{Thomas Jefferson National Accelerator Facility, Newport News, VA 23602, USA}
\affiliation{Hampton University, Hampton VA 23668, USA}

\date{\today}

\begin{abstract}
New nuclear structure function data from Jefferson Lab covering the higher $x$
and lower $Q^2$ regime make it possible to extract the higher order $F_2$
moments for iron and deuterium at low four--momentum transfer squared $Q^2$.
These moments allow for an experimental investigation of the nuclear momentum
sum rule and a direct comparison of the non-singlet nucleon moment with
Lattice QCD results.
\end{abstract}

%These are from moment analysis in proton paper
\pacs{13.60.Hb, 12.38.Qk, 25.30.Fj}

\maketitle

\section{Introduction}
Nuclear effects in lepton-nucleus scattering have been extensively studied,
both experimentally and theoretically, over the last few decades.  For recent
reviews, see Refs.~\cite{arneodo94, geesaman95}. The body of available data
provides clearcut evidence that the nucleus can not be simply described as a
collection of nucleons on mass shell. For example, the study of nuclear
structure functions led to the discovery of the ``EMC effect'' where it was
found that the quark distribution inside the nucleus differs from that of a
free nucleon. The availability of experimental information on the
$Q^2$ dependence of the moments of the nuclear structure function
$F_2^A(x,Q^2)$ has stimulated theoretical analyses of meson exchange
contributions and off--shell effects in nuclei, sometimes showing sizeable
deviations from predictions of simple convolution
models~\cite{benhar02,benhar00,cothran98,melnitchouk94}. In this discussion,
$A$ is the mass number, $Q^2$ is the four-momentum transfer squared in the
lepton-nucleon inclusive scattering process, and $x$ is the Bjorken scaling
variable, with $0<x<1$ for the proton, $0<x< M_A/M_p\approx A$ for a nucleus.

Previous nuclear structure function moment analyses have relied on moment
data extracted from several experiments carried out at CERN~\cite{aubert87,
aubert86} and SLAC~\cite{gomez94, arrington96} using $^{56}Fe$ and $^2H$
targets. The experimental values of Cornwall--Norton moments, $M_n(A,Q^2)$,
require precision measurements of structure functions covering large intervals
of $x, Q^2$, and $A$, since:
\begin{equation}
  \label{eq:defmom}
M_n(A,Q^2) = \int_0^A \; dx \; F_2^A(x,Q^2) \; x^{n-2}.
\end{equation}
Here, $n$ is an integer defining the order of the moments. We note that the
$n=2$ moment can be related to the familiar momentum sum rule, which must be
less than unity for the nucleon. Asymptotically, QCD predicts the fraction of
the nucleon momentum carried by the quarks to be $(1 + 16/3f)^{-1}$, where $f$
is the number of quark flavors~\cite{west85}.

Until recently, the set of experimental data at large $x$ was rather poor, and
thus the evaluation of the moments was correspondingly imprecise, especially
for large $n$. Typically, data were obtained in the deep inelastic scattering
regime at moderate to small values of $x$ and larger values of $Q^2$. One can
see immediately from Eq.~\ref{eq:defmom} that, as $n$ increases, larger $x$
data will increasingly dominate the moments. Additionally, at lower values of
$Q^2$, the structure function is larger in the higher $x$ region and dominates
even the lower order moments. Moreover, nuclear structure effects are expected
to show up most clearly at large values of $x$~\cite{kaptar93}.

Recently, data have become available from new experiments at Jefferson Lab
which cover higher $x$ and lower $Q^2$~\cite{arrington99, arrington01,
niculescu00a, niculescu00b}, complementing the previous data set. These new
data make it possible to accurately extract the moderate and lower $Q^2$
moments, and moments to higher orders. We report here results from a new
extraction of the $F_2$ structure function moments for iron and deuterium and
compare to proton data.

\section{Experiment}

\begin{figure}
\includegraphics[width=8cm,height=9cm,angle=0]{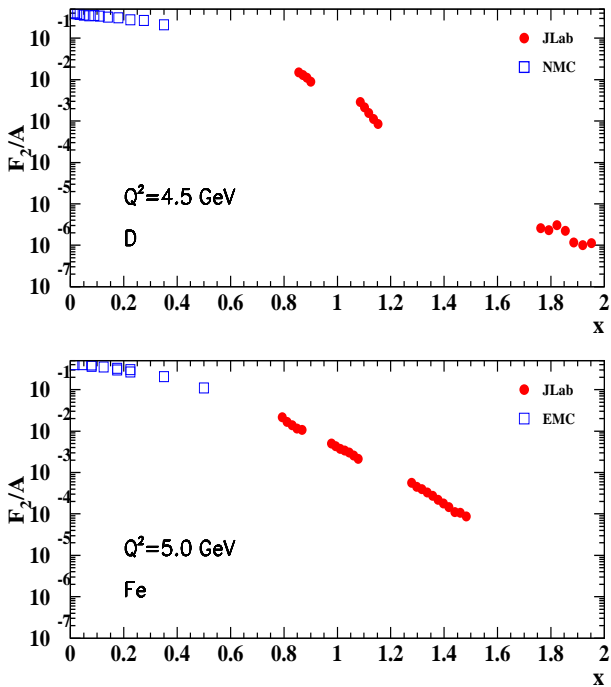}
\caption {(Color online) Example structure function data for deuterium (top)
and iron (bottom) at $Q^2=4.5$ GeV$^2$ and 5 GeV$^2$, respectively.}
\label{fig:fig1}
\end{figure}

Sample spectra used for the extraction of the moments are shown in
Fig.~\ref{fig:fig1} for deuterium at $Q^2=4.5$ and iron at 5~GeV$^2$. As noted
above, the calculation of the moment of a structure function requires data
covering the whole range in $x$ from 0 to $\approx A$ at a fixed $Q^2$. The
structure function data used in this analysis were obtained in experiments at
SLAC~\cite{dasu94, filippone92}, CERN~\cite{aubert86, berge91},
Fermilab~\cite{oltman92, adams96} and JLab~\cite{arrington99, arrington01,
niculescu00a, niculescu00b}. The $Q^2$ values where the best coverage in $x$
was available were selected. In some cases, the data were obtained not at
exactly the same $Q^2$ value. In these cases, a small range in $Q^2$, varying
from 0.01~GeV$^2$ at low $Q^2$ to 0.5~GeV$^2$ at high $Q^2$ was utilized. The
variations of the structure function over such ranges were smaller than 2\%.

As illustrated in Fig.~\ref{fig:fig1}, the data sets still do not cover the
full range in $x$, some extrapolations were necessary. Between data sets, two
methods were utilized, a spline fit and a simple linear extrapolation. Moments
obtained in such cases agreed within 2\%. To extrapolate to $x \to 0$ a
parameterization from NMC~\cite{arneodo95} was used for $Q^2>2$~GeV$^2$ and a
linear extrapolation was used for lower $Q^2$ data. The uncertainty introduced
by this extrapolation becomes negligible for higher order moments. The
extrapolation to $x \to A$, while negligible for $n=2$, becomes important for
the higher moments. The data used in this region were obtained at SLAC and
JLab and the coverage in $x$ is sufficient for most $Q^2$ and $n$ values. The
uncertainty in the moments due to the extrapolation to $x=A$ is less than 1\%
for $n=2$, around 3\% for $n=4$, up to 6\% for $n=6$, and up to 20\% for
$n=8$. The highest $x$ quasielastic and elastic contributions, important for
low $Q^2$, were calculated according to~\cite{dejager74,sargsian90} and added
to the moments.

In the near future, the extrapolations to $x\to 0$ and $x\to A$ can be improved
with new data coming from Jefferson Lab experiments~\cite{e99-118, e00-002,
e00-116, e02019, e02109, e03-103, e04-001}. These experiments have already
acquired data and results will become available over the next few years.
These newer data will allow for moments to be obtained over an expanded range
in $Q^2$, and for several additional nuclei, including $^3$He and $^4$He.

\section{RESULTS}

Tables~\ref{tab:d2_moments} and~\ref{tab:fe_moments} show the Cornwall--Norton
moments for deuterium and iron.  The uncertainties include published
experimental uncertainties on the structure functions, the uncertainties
due to the finite $Q^2$ range of the data and interpolation procedures,
extrapolations to low and high $x$, and the uncertainties in estimating
nuclear elastic and quasielastic contributions.  The combined uncertainties
are typically 5\%, except for low $Q^2$ values where the uncertainty in the
quasielastic become very large, especially for $n=2$.  At low $Q^2$, the
higher moments become increasingly dominated by the nuclear elastic
contribution, which is known to better than 5\%.  For the iron $n=6$ and
$n=8$ moments, the intermediate $Q^2$ values have large contributions from the
poorly known quasielastic contributions at extremely large $x$ values, and so
these moments are not included.

There are indications that two-photon exchange corrections to the
electron--nucleon elastic cross section might impact the extracted
moments~\cite{arrington04a}.  These corrections appear to be $\ltorder$6\% for
elastic $e$-$p$ scattering ($\ltorder$3\% for $e$-$n$~\cite{blunden05}),
peaking at large scattering angles. For the data included in this analysis, we
expect that two-photon exchange will contribute at most 2\% to the moments,
typically much less. This is small compared to the experimental uncertainties,
and these effects should partially cancel when comparing different nuclei.

\begin{table}
\begin{center}
\begin{tabular}{|c|c|c|c|c|}
\hline
  $Q^2$ & $n=2$   &    $n=4$         &   $n=6$  &    $n=8$   \\
(GeV$^2$)&    &           &   &      \\ \hline
  0.05  & .481$\pm$.481 & .807$\pm$.400 & 2.3618$\pm$.2362 &  8.5266$\pm$.8527 \\            
  0.10  & .407$\pm$.204 & .479$\pm$.120 & 1.0533$\pm$.0105 &  3.3723$\pm$.3372 \\                 
  0.20  & .320$\pm$.080 & .284$\pm$.034 & 0.3946$\pm$.0395 &  0.7653$\pm$.0765 \\                 
  0.45  & .296$\pm$.021 & .193$\pm$.019 & 0.2163$\pm$.0216 &  0.2968$\pm$.0359 \\                 
  0.80  & .220$\pm$.011 & .092$\pm$.005 & 0.0844$\pm$.0060 &  0.0961$\pm$.0103 \\                 
  1.50  & .180$\pm$.009 & .040$\pm$.003 & 0.0261$\pm$.0020 &  0.0235$\pm$.0033 \\                 
  2.40  & .169$\pm$.008 & .028$\pm$.001 & 0.0165$\pm$.0010 &  0.0156$\pm$.0008 \\                 
  3.20  & .162$\pm$.008 & .021$\pm$.001 & 0.0091$\pm$.0005 &  0.0065$\pm$.0003 \\                 
  4.50  & .165$\pm$.008 & .016$\pm$.001 & 0.0056$\pm$.0003 &  0.0039$\pm$.0002 \\             
  5.00  & .161$\pm$.008 & .017$\pm$.001 & 0.0052$\pm$.0003 &  0.0030$\pm$.0002 \\                 
  7.00  & .163$\pm$.008 & .016$\pm$.001 & 0.0038$\pm$.0002 &  0.0015$\pm$.0001 \\
\hline
\end{tabular}
\caption{Moments of the $F_2$ structure function per nucleon for the deuteron.}
\label{tab:d2_moments}
\end{center}
\end{table}

\begin{table}
\begin{center}
\begin{tabular}{|c|c|c|c|c|}
\hline
  $Q^2$ & $n=2$   &    $n=4$         &   $n=6$  &    $n=8$   \\
(GeV$^2$)& & & & \\ \hline
  0.05  & .203$\pm$.203 & 204$\pm$10    & (6.4$\pm$.32)$\times$10$^5$ &  (2.0$\pm$.1)$\times$10$^9$ \\
  0.10  & .207$\pm$.100 & 5.74$\pm$.289 & (1.77$\pm$.09)$\times$10$^4$&  (5.6$\pm$.28)$\times$10$^7$ \\
  0.25  & .277$\pm$.069 & .273$\pm$.137 & 2.763$\pm$1.242   &  6600$\pm$330 \\                 
  0.40  & .265$\pm$.027 & .273$\pm$.041 & ---               &  --- \\                 
  1.00  & .209$\pm$.010 & .095$\pm$.005 & 0.276$\pm$0.044   &  --- \\                 
  1.90  & .166$\pm$.008 & .034$\pm$.002 & 0.0270$\pm$0.0015 &  .0447$\pm$.0058 \\                 
  2.90  & .174$\pm$.009 & .018$\pm$.001 & 0.0114$\pm$0.0010 &  .0146$\pm$.0063 \\                 
  5.00  & .158$\pm$.008 & .015$\pm$.001 & 0.0050$\pm$0.0004 &  .0032$\pm$.0006 \\                 
  6.00  & .164$\pm$.008 & .016$\pm$.001 & 0.0038$\pm$0.0002 &  .0020$\pm$.0004 \\
\hline
\end{tabular}
\caption{Moments of the $F_2$ structure function per nucleon for iron.}
\label{tab:fe_moments}
\end{center}
\end{table}

The lower $n$ moments display a very shallow to negligible $Q^2$ dependence.
In the Operator Product Expansion (OPE), higher twist effects (interactions
between the struck quark and other quarks in the electron--nucleon scattering
process) are expected to manifest a $1/Q^2$ dependence in the moment. This is
not observed in the data, which is somewhat surprising at these low $Q^2$
values where such effects could be large. The asymptotic behavior of the
second moment is ultimately governed by the energy--momentum tensor in the OPE
and, thus, has no $Q^2$ dependence, as in the quark--parton model~\cite{west85}.
Even at the low $Q^2$ values studied here, the moments display this
quark--parton model behavior over most of the $Q^2$ range. The lower $Q^2$
moments are dominated by high $x$ resonance regime. Hence, this observation is
yet another striking manifestation of quark--hadron duality~\cite{wally2}.

The higher $n$ moments, on the other hand, do display an increased $Q^2$
dependence. These data may therefore be used for precision higher twist
extractions. However, the higher $n$ moments are increasingly dominated by the
high $x$, including the elastic and quasi--elastic regimes, where the $x$ and
$Q^2$ dependences are less well understood in terms of the OPE.
 
If nuclear effects are small, the moments for iron can also be constructed by
adding the proton and neutron contributions, extracted from
proton~\cite{armstrong01} and deuteron data. To investigate how well this
simplified approach works, the following simple formula was employed:
\begin{equation}
M_n(Fe)=Z\times M_n(p) + (A-Z)\times M_n(n),\label{eq:sum_rule}
\end{equation}
\noindent where $M_n(n)$ is taken to be $M_n(d)-M_n(p)$. Here, $M_n(p),
M_n(n)$, and $M_n(d)$ refer to the $n$th moment of the proton, neutron, and
deuteron, respectively, and $Z$ is the atomic number of iron. This is equivalent to
extracting the iron data as 28 deuterons with a small neutron excess
contribution. Simple Fermi motion should not yield a significant nuclear
dependence in the $M_2$ moment, and off-shell effects have been
studied~\cite{melnitchouk94, cothran98} and are also expected to be small for
the lowest moment, and on the order of 10\% for moments up to
$n=5$~\cite{cothran98}.

\begin{figure}
\includegraphics[width=8.5cm,angle=0]{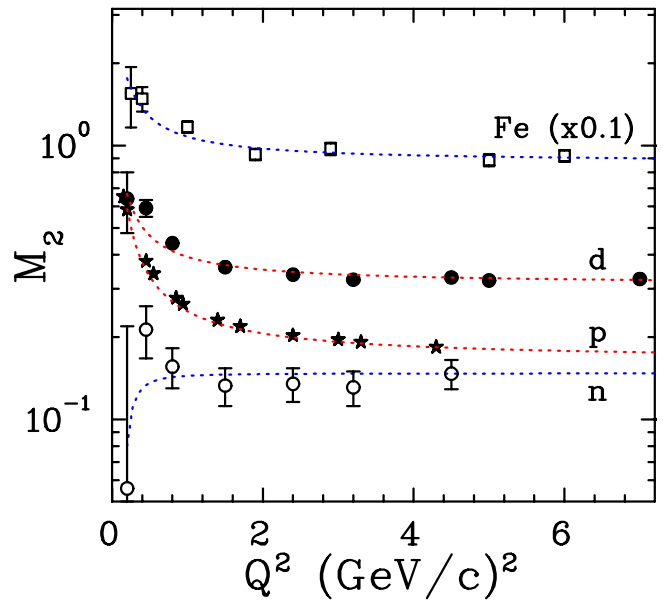}
\caption {(Color online) The second moment of $F_2$ for proton
(stars), deuteron (full circles), and iron (squares).  The hollow
circles are the neutron moments taken from the difference of deuteron
and proton. The red dashed lines are fits to the deuteron and proton moments,
and the blue dashed lines are the neutron and iron moments extracted from
these fits using the procedure described in the text.}
\label{fig:energy_sum_rule}
\end{figure}

This procedure is illustrated in Fig.~\ref{fig:energy_sum_rule} for the second
moment, $M_2$. The iron data are shown as squares, deuteron data as full
circles, proton data as stars. The red dashed lines describing the deuteron
and the proton moments are simple fits to the proton and deuteron data, which
are then used to calculate the neutron moment, $M_2(d)-M_2(p)$, and the iron
moment as 26 protons and 30 neutrons, as described in Eq.~\ref{eq:sum_rule}. 
No additional correction was made for nuclear effects or the non--isoscalarity
of the target. The neutron and iron moments thus calculated are shown as blue
dashed lines in the figure, while the hollow circles show the neutron moments
taken directly from the difference of deuteron and proton moments. For $Q^2 >
4$~GeV$^2$, the ratio of $M_2(Fe) / M_2(D)$, normalized to the number of
nucleons, is $0.99\pm0.05$, consistent with the value 0.96~\cite{liuti_priv},
from a calculation based on Ref.~\cite{cothran98}.

\begin{figure}
\includegraphics[width=8cm,angle=0]{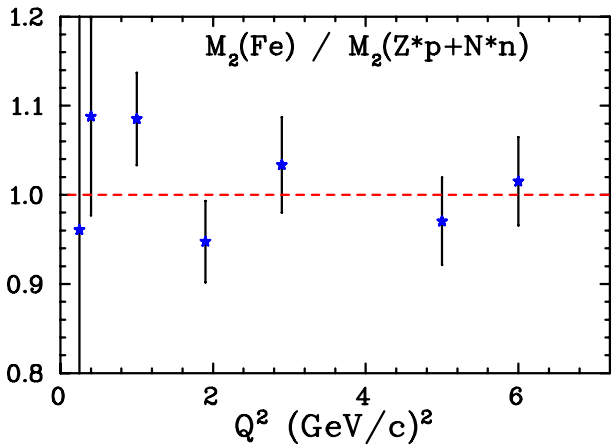}
\caption {(Color online) The ratio of the QCD moments for iron calculated
using iron data to the moments constructed using deuterium and proton data
shown as a function of $Q^2$.} \label{fig:fig3}
\end{figure}

The ratios of the measured moments for iron compared to the moments taken from
the deuteron and proton moments are shown in Fig.~\ref{fig:fig3}. It can be
seen that these two methods yield the same results within the uncertainty.
Combining all of the values yields a deviation of ($0.9 \pm 2.2$)\%, or
($0.5\pm2.9$)\% if we consider only $Q^2>2.5$~GeV$^2$. This result contradicts
interpretations of the EMC effect that predict significant modification to the
the {\it total} quark momentum distribution in nuclei.  However, it is
consistent with other interpretations where the total quark momentum is
conserved~\cite{wally1, geesaman95}.  Here, the data indicate that the
integrated iron nucleus can be described well as simply being composed of free
deuterons, with a minimal correction for neutron excess in $26 p + 30 n$. It
seems the EMC effect is a redistribution of quark momentum without any
additional momentum added by the nuclear environment outside of whatever is
already present in the deuteron.

One can also connect the nuclear dependence of the quark distributions
to the coordinate space parton distributions~\cite{hoyer96, vanttinen98}.
The $A$ dependence of the $n=2$ moment is then related to the $A$ dependence
of the light cone distributions at short distance.  The fact that the data
indicate extremely small nuclear effects is consistent with the result that
the $A$ dependence for distances less than the inter-nucleon spacing is
surprisingly small ($<$2\%), due to cancellation between the shadowing,
anti-shadowing, and EMC regions~\cite{hoyer96}.

We note, further, that the redistribution can be quite large, locally. In the
structure functions at fixed $(x,Q^2)$ values, there are drastic differences
in the nucleon and in nuclei. For instance, resonance structure can be
observed in the $\Delta$ resonance region in deuterium but not at all in iron.
 However, the effect of this redistribution is smaller in nuclei, such that
the resonance region structure function nearly reproduces the DIS structure
function in nuclei~\cite{filippone92, arrington01}, and the ratios of the
nuclear structure function in the resonance region reproduce the observed EMC
effect with high precision~\cite{arrington03c}.

West points out the need to reconcile the difference between the
fundamental asymptotic QCD sum rule,
\begin{equation}
\int_0^A ({1\over A} F_2^A - {1\over 2}F_2^D) dx = 0,
\label{eq:west1}
\end{equation} 
\noindent based on energy--momentum conservation, and the nominal observation
of the EMC effect that the nuclear structure function is not simply $A$ times
that of a nucleon~\cite{west85}. The new data presented here
(Fig.~\ref{fig:fig3}) indicate agreement with this sum rule already at the low
$Q^2$ values here observed. As a quantitative example, the integral in
Eq.~\ref{eq:west1} becomes of the order of $4\times 10^{-3}$ compared to
individual moments of $\sim$0.2 at $Q^2=2.9$~GeV$^2$.

The moments of the structure function $F_2$ can be determined theoretically on
the lattice~\cite{detmold02, dolgov02, gockeler04}. While contributions from
disconnected diagrams~\cite{dolgov02} make it more difficult to calculate the
separate proton and neutron moments on the lattice, these contributions cancel
in the non-singlet combination $M_n(p)-M_n(n)$. To compare our results with
lattice calculations we extracted the difference between the proton and
neutron moments for $n$$=$2 and $n$$=$4. We assume that the deuteron moment is
equal to the sum of proton and neutron, and then determining the $p-n$ moment
from the proton~\cite{armstrong01} and deuteron moments, taking $M_n(p-n) =
2M_n(p) - M_n(d)$. Because the proton and deuteron moments are sometimes
extracted at slightly different $Q^2$ values, we combine our extracted
deuteron moments with the nearest proton moments, scaling the proton to the
correct $Q^2$ value using the $Q^2$ dependence of the simple fit shown in
Fig.~\ref{fig:energy_sum_rule}. The extracted values for the $M_2$ and $M_4$
moments for $p-n$ are independent of $Q^2$ above 2~GeV$^2$. The experimental
results shown in Table~\ref{tab:lattice1} for $Q^2 \approx 4$~GeV$^2$ come
from combining the extracted values at $Q^2=3.2$ and 4.5~GeV$^2$, and are
compared to lattice calculation at $Q^2=4$~GeV$^2$. Because the proton and
neutron $n=2$ moments are comparable in size, there is a large cancellation in
the difference which leads to the large relative uncertainty.

The $n=2$ moment from Detmold \etal~\cite{detmold02} is in excellent agreement
with the measured data. For $n=4$, the small discrepancy between the lattice
calculation and our experimental result could be due to higher twist effects,
which are not included in the lattice result, although the $Q^2$ dependence of
the moments does not indicate that these are large. In addition, no nuclear
effects were taken into consideration when extracting the neutron moment from
deuterium data. These effects seem to be small when averaged over the entire
$x$ range but they might still have some non-negligible contribution. It
should also be noted that there are still open issues for lattice
calculations, such as chiral extrapolation, volume dependence,
or renormalization. To demonstrate this, we also show the results of Dolgov
\etal~\cite{dolgov02}, and Gockeler \etal~\cite{gockeler04}. The main
difference between the lattice calculations presented here is the chiral
extrapolations used.  In Ref.~\cite{dolgov02}, the lattice results are
extrapolated linearly to the physical limit, while in Ref.~\cite{detmold02},
the extrapolation includes the correct chiral behavior from chiral effective
theory.

\begin{table}
\begin{center}
\begin{tabular}{|c|c|c|c|c|}
\hline
        & This work        &  Detmold &   Dolgov & Gockeler \\
        & $Q^2 \approx 4$~GeV$^2$ & \etal~\cite{detmold02}  & \etal~\cite{dolgov02} & \etal~\cite{gockeler04} \\ \hline
  $n=2$ & 0.049(17)  & ~0.059(8)~    & 0.269 & 0.245 \\            
  $n=4$ & 0.015(03)  &  0.008(3)     & 0.078 & 0.059 \\                 
\hline
\end{tabular}
\caption{Moments of the $F_2$ structure function for the difference
        $p-n$. Experimental results for $Q^2 \approx 4$~GeV$^2$ from the
        present work are compared to lattice calculations at 4 GeV$^2$.}
\label{tab:lattice1}
\end{center}
\end{table}

We note that comparisons between lattice and nominal data formed from
pdf--based fits have been performed previously~\cite{detmold02}. We stress
that such fits do not adequately account for the large $x$ regime where
they are unconstrained by data. Moreover, substantial uncertainties exist in
the down quark distribution, $d(x)$, associated with assumptions utilized in
extracting neutron results from deuteron data, as well as the unknown behavior
of $d/u$ as $x \to 1$.

\section{Conclusions}

In conclusion, we utilized inclusive electron--nucleus scattering data to
obtain nuclear structure function moments for iron and deuterium. The new data
are particularly important for moment calculations at low $Q^2$, where there
was a paucity of previous data. Moreover, at low $Q^2$ and higher $n$, the
need for large $x$ data increases as this regime comes to dominate the moments.

Negligible $Q^2$ dependence is observed in the lower order moments, indicating
agreement with asymptotic predictions and minimal higher twist effects. This
is surprising, given that the data extend to quite low $Q^2$ values.

The $n=2$ moment, related to the momentum sum rule, is here presented for both
iron and deuterium. Additionally, a neutron momentum was formed by subtracting
existing proton data from the deuterium data. The measured iron moments were
found to agree with moments simply constructed from these neutrons and
protons. This observation has interesting implications for interpretations of
the EMC effect.

Finally, these neutron and proton moment data allow for comparison with
lattice QCD calculations. The extracted non-singlet moments provide the first
direct comparison to lattice calculations of the non-singlet moments and the
results are in good agreement with the calculation of Ref.~\cite{detmold02}.

\begin{acknowledgments}

This work was supported in part by DOE Grants W-31-109-ENG-38 and
DE-FG02-95ER40901, and NSF Grants 0099540 and 0245045.
We thank Wally Melnitchouk for useful discussions.
\end{acknowledgments}

\bibliography{nuclear_moments_2005}

\end{document}